\documentclass[a4paper,11pt]{article}
\usepackage{pos}

\usepackage{subfig}

\usepackage{xcolor}
\usepackage{todonotes}

\title{Searching for neutrino transients below 1 TeV with IceCube}

\author{The IceCube Collaboration \\{\normalsize \normalfont(a complete list of authors can be found at the end of the proceedings)}}

\emailAdd{mlarson@icecube.wisc.edu}

\abstract{Recent observations of GeV gamma-rays from novae have led to a paradigm shift in the understanding of these objects. While it is now believed that shocks contribute significantly to the energy budget of novae, it is still unknown if the emission is hadronic or leptonic in origin. Neutrinos could hold the key to definitively differentiating between these two scenarios, though the energies of such particles would be much lower than are typically targeted with neutrino telescopes. IceCube's densely instrumented DeepCore sub-array provides the ability to reduce the threshold for observation from 1 TeV down to approximately 10 GeV. We will discuss recent measurements in this low energy regime, details of a new sub-TeV selection, and prospects for future searches for transient neutrino emission.

\vspace{4mm}
{\bfseries Corresponding authors:}
Michael Larson$^{1,*}$, Jason Koskinen$^{2}$, Alex Pizzuto$^{3}$, Justin Vandenbroucke$^{3}$

{$^{1}$ \itshape Dept. of Physics, University of Maryland, College Park, MD 20742, USA}\\
{$^{2}$ \itshape Niels Bohr Institute, University of Copenhagen, DK-2100 Copenhagen, Denmark}\\
{$^{3}$ \itshape Dept. of Physics and Wisconsin IceCube Particle Astrophysics Center, University of Wisconsin{\textendash}Madison, Madison, WI 53706, USA}\\[4mm]
$^*$ Presenter
\FullConference{37$^{\rm{th}}$ International Cosmic Ray Conference (ICRC 2021)\\
		July 12th -- 23rd, 2021\\
		Online -- Berlin, Germany}
}

\begin{document}
\maketitle

\section{Introduction}
\label{sec:introduction}

In 2013, the IceCube Collaboration announced the first discovery of a diffuse extraterrestrial flux of neutrino events\cite{Aartsen:2013jdh}. 
Since that initial discovery, many searches for the source of the diffuse flux of neutrinos have been performed, both by the IceCube collaboration and by the wider community.
To date, few potential sources have shown an excess of events in IceCube analyses with the notable exception of TXS~0506+056\cite{IceCube:2018cha}.
While searches continue, more than ten years of data have already been collected and increased integration time can only slowly improve existing limits.
New methods or samples provide a viable way to significantly improve the physics reach of the existing IceCube detector.

In most searches for sources of astrophysical neutrino emission, charged-current muon neutrino interactions with energies of TeV or higher are used. 
These events produce long-ranging muons visible in the IceCube detector that can be reconstructed to within 1$^\circ$ of their source. 
Backgrounds from muons produced in cosmic ray induced air showers above the detector provide the dominant background in the southern sky, severely limiting searches for astrophysical sources.
In order to better detect sources in the southern sky, neutrinos of other flavors and interaction types are necessary.

New samples are under development using cascade-like events\cite{Sclafani:2021icrc} and starting track events\cite{Silva:2021icrc}.
These samples are sensitive to the same TeV emission energy range as the through-going track-like analyses typically performed by IceCube, but can use vetoing techniques to reduce the background from atmospheric muons, leading to significant improvements in physics reach in the southern sky.
Neutrino events at energies below a few hundred GeV are largely unexplored, however.
By utilizing and improving upon samples designed for atmospheric oscillations research, IceCube can test for astrophysical neutrino sources in this energy regime. 

\section{Transients with Three Years of GRECO}
\label{sec:mia_analysis}
IceCube is a cubic-kilometer neutrino detector installed in the ice at the geographic South Pole\cite{Aartsen:2016nxy} at depths of 1450 m and 2450 m below the surface.
A total of 5160 digital optical modules (DOMs) are arranged on 86 strings.
Reconstruction of the direction, energy and flavor of the neutrinos relies on the optical detection of Cherenkov radiation emitted by charged particles produced in the interactions of neutrinos in the surrounding ice or the nearby bedrock.
The DeepCore subarray as defined in this analysis includes 8 densely instrumented strings optimized for low energies plus 12 adjacent standard strings.
DeepCore lowers the energy threshold of the IceCube detector from 100 GeV to about 5 GeV.

Event samples using DeepCore have been used in several analyses, including searches for neutrino oscillations\cite{Aartsen:2019tjl}.
DeepCore has also occasionally been used for astrophysical searches\cite{Aartsen:2015eai}, although only rarely.
Using vetoing techniques\cite{Collaboration:2011ym}, atmospheric muons from the southern sky can be reduced while maintaining large numbers of atmospheric neutrinos.

The GRECO (GeV-Reconstructed Events with Containment for Oscillations) oscillation event selection (analysis $\mathcal{A}$ of \cite{Aartsen:2019tjl}), originally developed for measurement of tau neutrino appearance, consists of data taken between April 2012 and May 2015, and may be used to search for low energy astrophysical neutrino emission. 
The GRECO oscillation sample has a total rate of 0.87 mHz, primarily due to atmospheric muon neutrino interactions at energies around 20 GeV. 
Events are reconstructed with a hypothesis assuming both a hadronic interaction and an outgoing muon track.
The muon track is assumed to be minimum ionizing and the track length is included as a parameter in the reconstruction, allowing a smooth transition between the standard cascade and track fits used in IceCube. 
Most events used for higher energy searches are through-going tracks with average directional reconstruction errors of less than 1$^\circ$.
DeepCore events are low energy starting tracks with approximately 10-20$^\circ$ resolution or cascades with resolutions of about 30-40$^\circ$.

\begin{figure}
    \centering
    \includegraphics[width=0.85\textwidth]{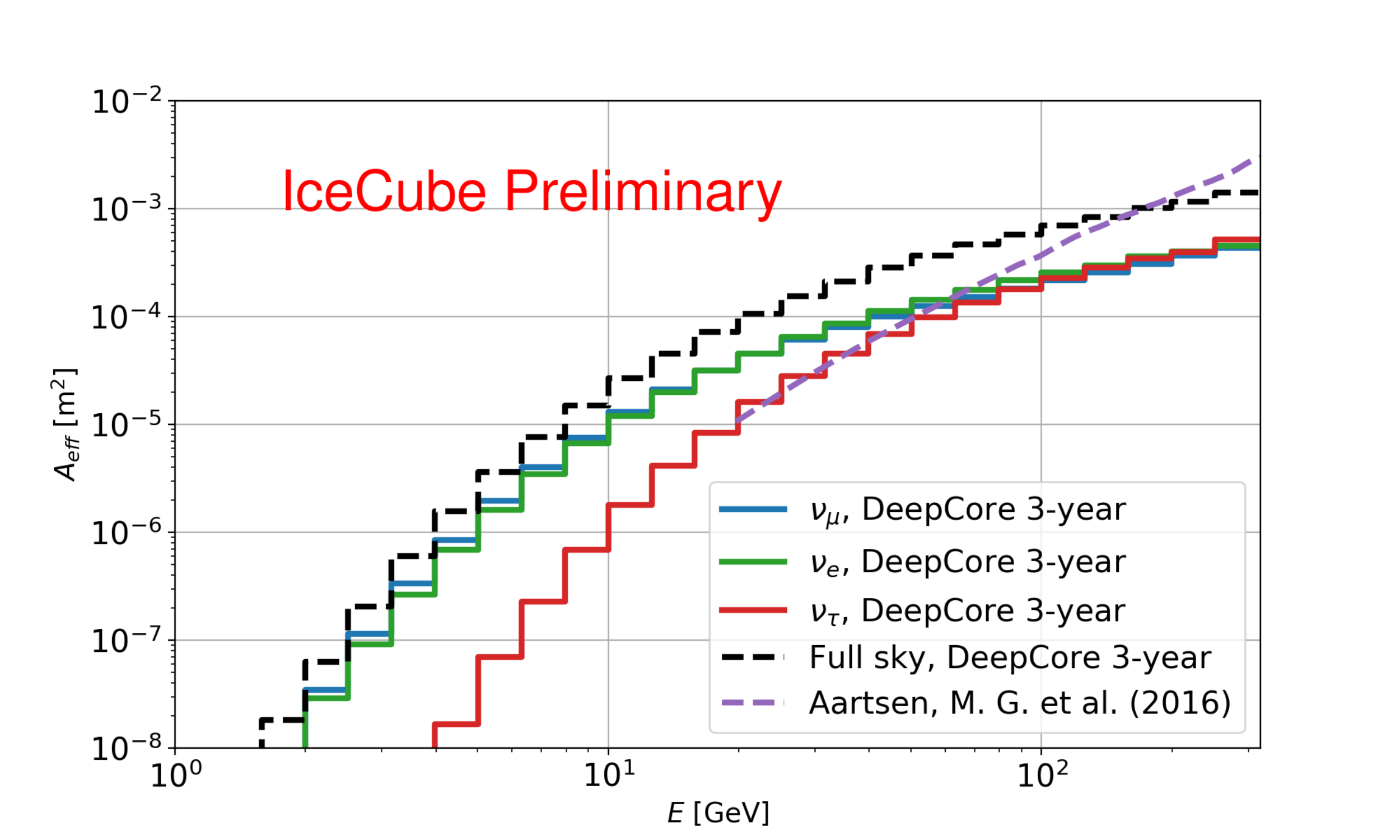}
    \caption{The effective area for each of the three neutrino flavors in the three-year GRECO oscillation sample compared to a previous DeepCore-based astrophysical neutrino sample\cite{Aartsen:2015eai}. Significant improvements in total effective area of the GRECO oscillation sample (black, dashed) compared to the previous analysis (purple, dashed) are visible below 100 GeV.}
    \label{fig:greco_osc_aeff}
\end{figure}

The GRECO oscillation sample provides roughly uniform effective area, shown in Figure~\ref{fig:greco_osc_aeff}, across the sky and across both muon and electron neutrinos below 100 GeV and shows a significant improvement relative to previous DeepCore analyses.
The large atmospheric flux and large angular uncertainties provide a significant barrier for astrophysical neutrino searches, however.
To limit the contributions from background, the GRECO oscillation sample is used to search for transient neutrino flares of timescales less than 600 seconds, giving an average expectation of 0.5 atmospheric background events per flare across the full sky. 

An untriggered flare search has been performed using the GRECO oscillation sample\cite{Abbasi:2020ddb}. 
Time periods with potential flares are identified by representing each event with a Gaussian kernel

\begin{equation}
    K(t; t_i, \Delta) = \frac{1}{\Delta \sqrt{2\pi}} \mathrm{exp}\left(\frac{-\left(t-t_i\right)^2}{2\Delta^2}\right)
\end{equation}
where $t_n$ is the time of the observed event and $\Delta$ is an assumed bandwidth of 100~s. 
A kernel density estimate (KDE) is produced by summing the contributions of each Gaussian kernel for each time $t$

\begin{equation}
    R_{KDE}(t) = \sum_{i}^{N} K(t; t_n, \Delta)
\end{equation}
where $N$ is the total number of events.
Periods of interest are defined by searching for periods when $R_{KDE}(t)$ is above a threshold of 905.80, chosen to limit the expected number of background search windows to 100 across the available three years.
The center of each period above threshold is used to define the center of a search window of 600~s.

An unbinned likelihood is applied to each search window, given by

\begin{equation}
\label{eq:likelihood}
    \mathcal{L}(n_s) = \frac{\left(n_s+\left<n_b\right>\right)^N}{N!}e^{-\left(n_s+\left<n_b\right>\right)}\prod_{i}^{N}\left(\frac{n_s \mathcal{S}_i}{n_s+\left<n_b\right>} + \frac{\left<n_b\right> \mathcal{B}_i}{n_s+\left<n_b\right>}\right),
\end{equation}
where $n_s$ is the number of signal events, $\left<n_b\right>$ is the expected number of background events in the search window, $N$ is the total number of events observed, and $\mathcal{S}_i$ and $\mathcal{B}_i$ are signal and background probability density functions (PDFs) for each event $i$.
A likelihood ratio is used a as a test statistic

\begin{equation}
    TS = \log\frac{\mathcal{L}(\hat{n}_s)}{\mathcal{L}(0)} = -\hat{n}_s + \sum_i^N\log\left(\frac{\hat{n}_s \mathcal{S}_i}{\left<n_b\right>\mathcal{B}_i}+1\right)
\end{equation}
with $\hat{n}_s$ as the best-fit number of signal events observed during minimization.

The background PDF $\mathcal{B}_i$ is assumed to be constant in right ascension, but can vary as a function of declination. 
The PDF is built using experimentally observed data assuming that signal contributions are negligible relative to atmospheric backgrounds.
Events are binned in 25 bins of sin($\delta$) and splined to provide a smooth distribution.

The signal PDF $\mathcal{S}_i$ assumes a Kent distribution with circularized errors of the form

\begin{equation}
    \mathcal{S}_i = \frac{\kappa_i}{4\pi \sinh\left(\kappa_i\right)} \mathrm{exp}\left(\kappa_i \cos\left(\left|\vec{x}_{source}-\vec{x}_i\right|\right)\right)
\end{equation}
where $\left|\vec{x}_{source}-\vec{x}_i\right|$ is the angular distance between the assumed source position and event $i$ and $\kappa_i$ is related to the estimated angular resolution $\sigma_i$ by 

\begin{equation}
    \kappa_i \approx \frac{1}{\sigma_i^2}.
\end{equation}

The values of $\sigma_i$ are estimated using the median angular error as a function of energy and declination calculated from simulated signal spectra.
Estimates for $\sigma_i$, shown in Figure~\ref{fig:greco_astro_resol}, are performed separately for "track-like" (reconstructed track length longer than 50~m) or "cascade-like" (reconstructed track length shorter than 50~m) events to exploit differences in angular resolutions between the two event classes.

\begin{figure}
    \centering
    \subfloat{\includegraphics[width=0.4\textwidth]{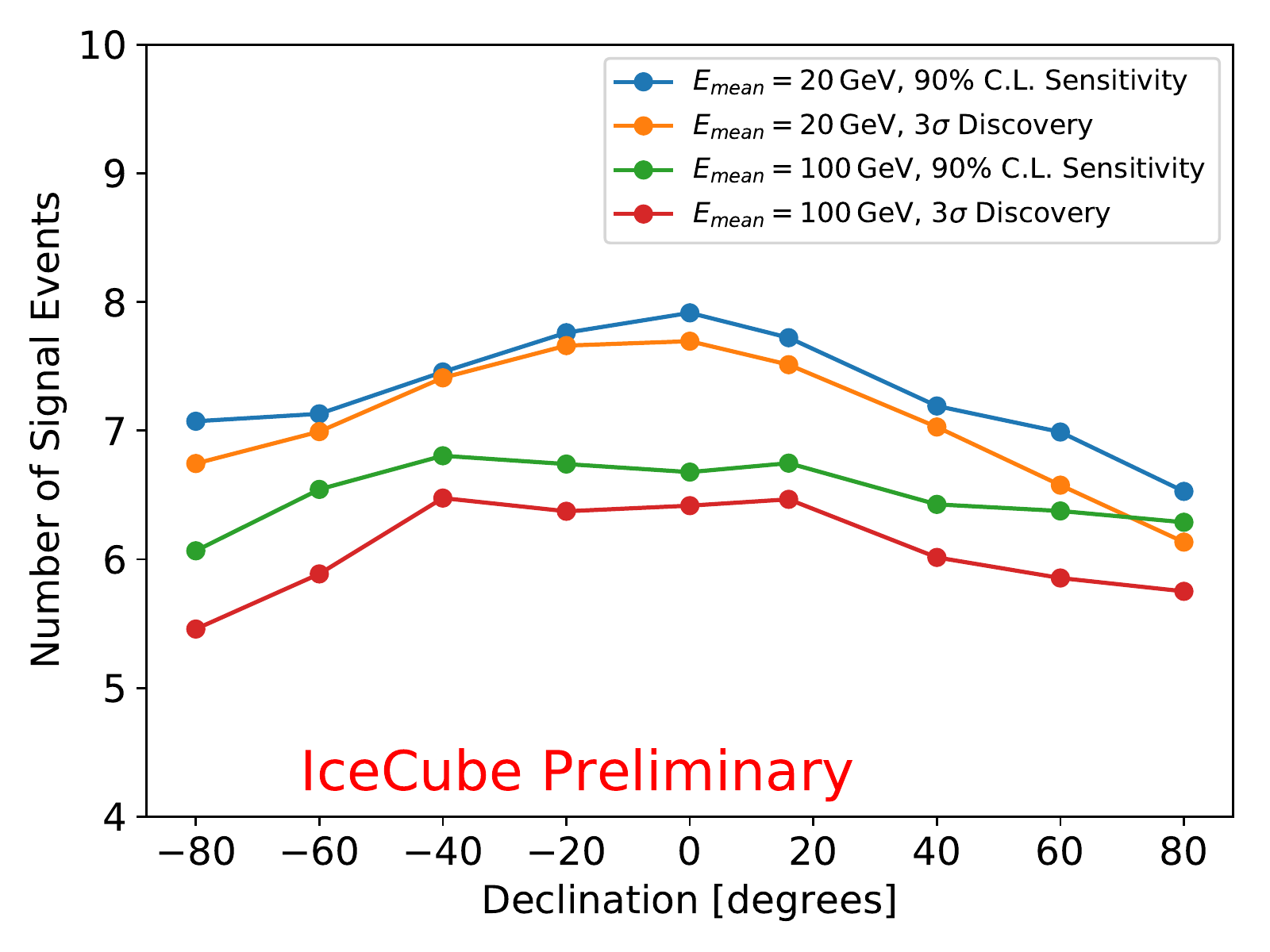}}
    \qquad
    \subfloat{\includegraphics[width=0.4\textwidth]{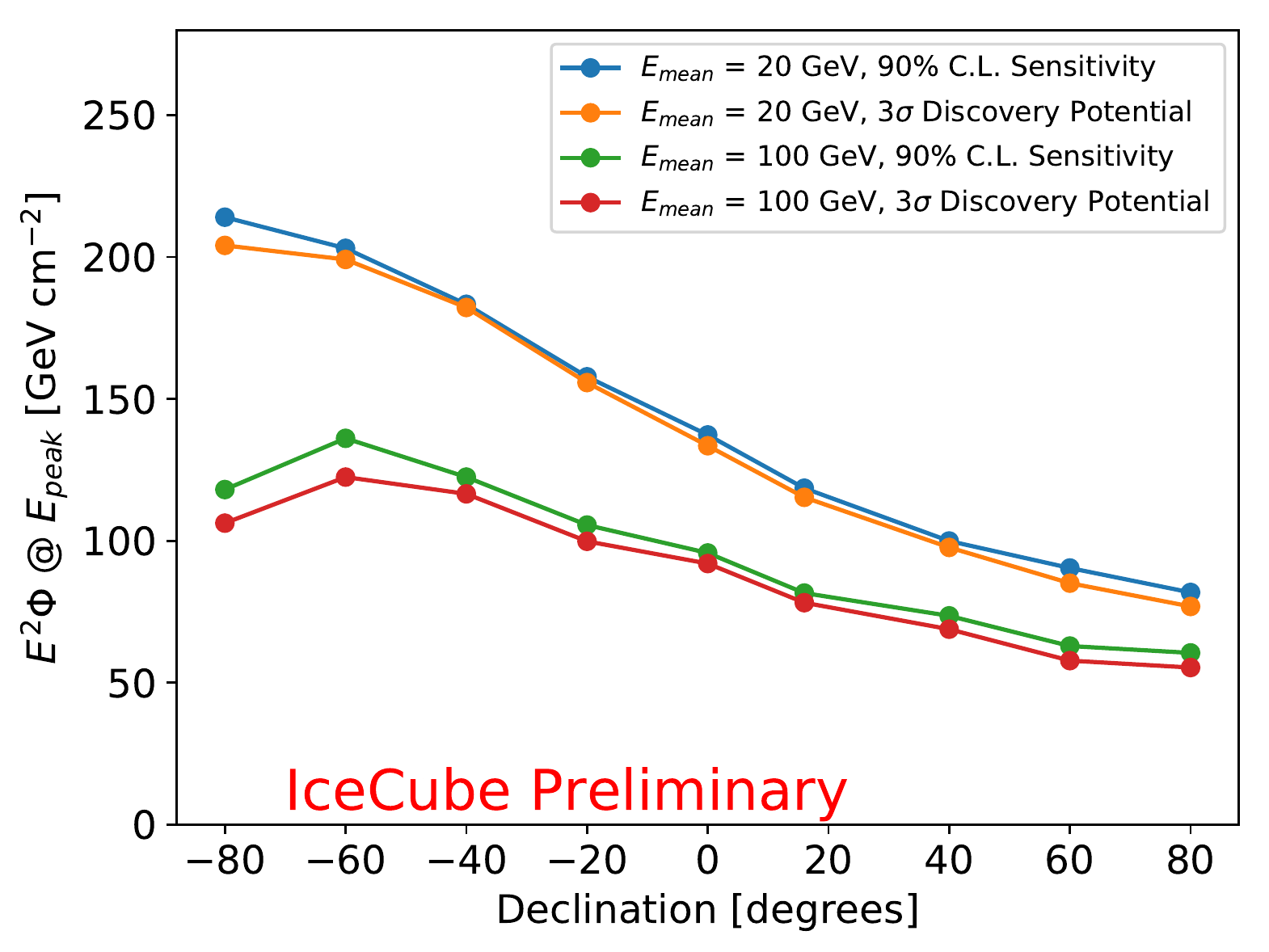}}
    \caption{Sensitivity (left) and discovery potential (right) for the GRECO oscillation sample using the Dirac flux in Equation~\ref{equation:dirac_spectrum} assuming $E_{mean}$ of 20~GeV and 100~GeV. }
    \label{fig:greco_osc_sensitivity}
\end{figure}

The sensitivity of the analysis is evaluated for the GRECO oscillations sample by assuming a spectrum\cite{Murase:2013hh} of the form

\begin{equation}
    \label{equation:dirac_spectrum}
    \Phi\left(E\right) = \Phi_0 \left(\frac{E}{E_{\mathrm{mean}}/3.15}\right)^2 / \left(e^\frac{E}{E_{\mathrm{mean}}/3.15}+1\right).
\end{equation}

Simulated background-only measurements are performed on data scrambled in right ascension processed with the KDE methods described above to identify search windows. 
Signal simulations include data scrambled in right ascension and signal events from sources generated randomly across the sky. Signal flare times are drawn from a uniform distribution covering the event sample. 
Simulated flares are produced with a Gaussian time profile with a width of 100~s.
The 90\% sensitivity - defined as the median expected 90\% upper limit - and 3$\sigma$ discovery potential - the median flux required for a positive result at 3$\sigma$ significance - of the analysis are shown in Figure~\ref{fig:greco_osc_sensitivity}.

A total of 300 search windows were identified in the unblinded dataset with 267 search windows containing more than one estimated signal event. 
After accounting for trials, the final result is consistent with background fluctuations, with p>0.5.

\section{The Realtime GRECO Selection}
\label{sec:greco_online}

\begin{figure}
    \centering
    \includegraphics[width=0.5\textwidth]{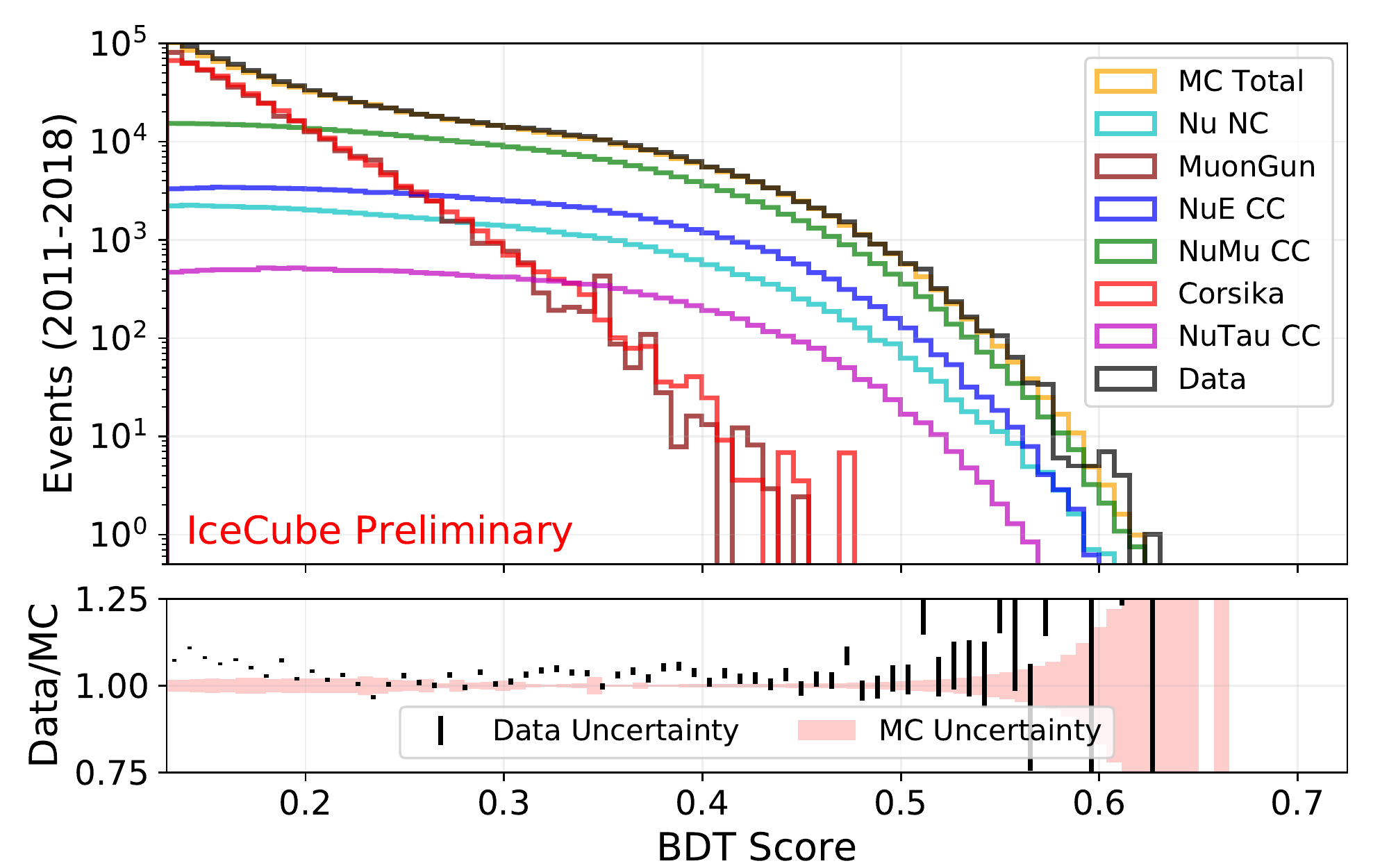}
    \caption{The expected contributions from each atmospheric background component, including neutrinos and two independent simulations of atmospheric muons (labeled CORSIKA and MuonGun). No fitting of atmospheric components has been performed. Events with a BDT score below 0.13 are removed, yielding a sample with 60\% atmospheric neutrinos and 40\% atmospheric muons.}
    \label{fig:greco_astro_bdt}
\end{figure}
The merging of two independent boosted decision trees (BDTs) in the GRECO oscillation sample and reoptimization of several variables can provide significant improvements in effective area for astrophysical searches.
An updated version of the selection has been developed to include these improvements. 
An updated version of the GENIE neutrino generator\cite{Andreopoulos:2009rq}, new charge calibrations in both data and simulation, an updated model of the glacial ice properties, and newly available high statistics simulation sets of atmospheric muons are used to train a single BDT using a modified set of the GRECO oscillation variables\cite{Aartsen:2019tjl}. 
The results from the new BDT training, given in Figure~\ref{fig:greco_astro_bdt}, show that the data is well-modeled by atmospheric-weighted simulation. 

\begin{figure}
    \centering
    \subfloat{\includegraphics[width=0.5\textwidth]{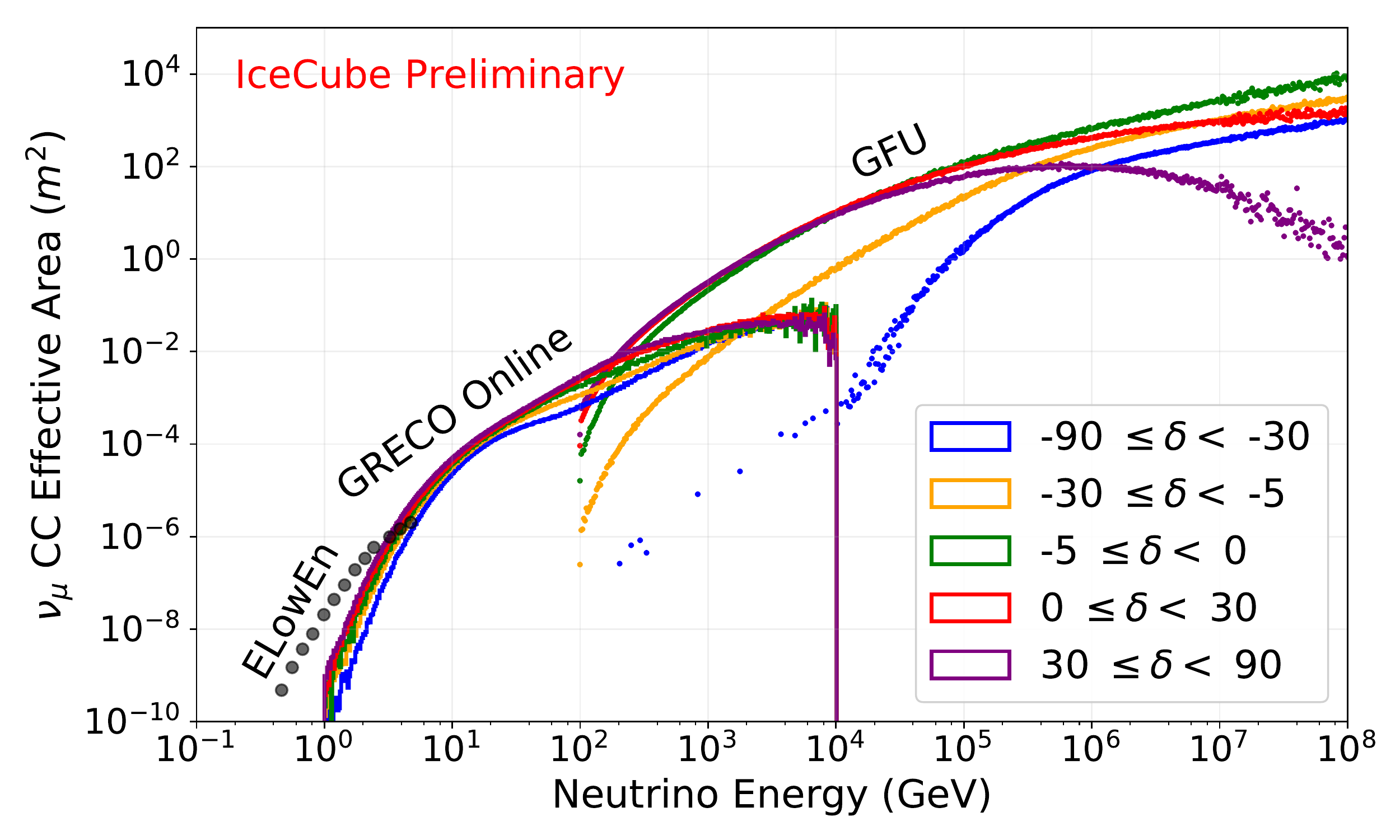}}
    \subfloat{\includegraphics[width=0.5\textwidth]{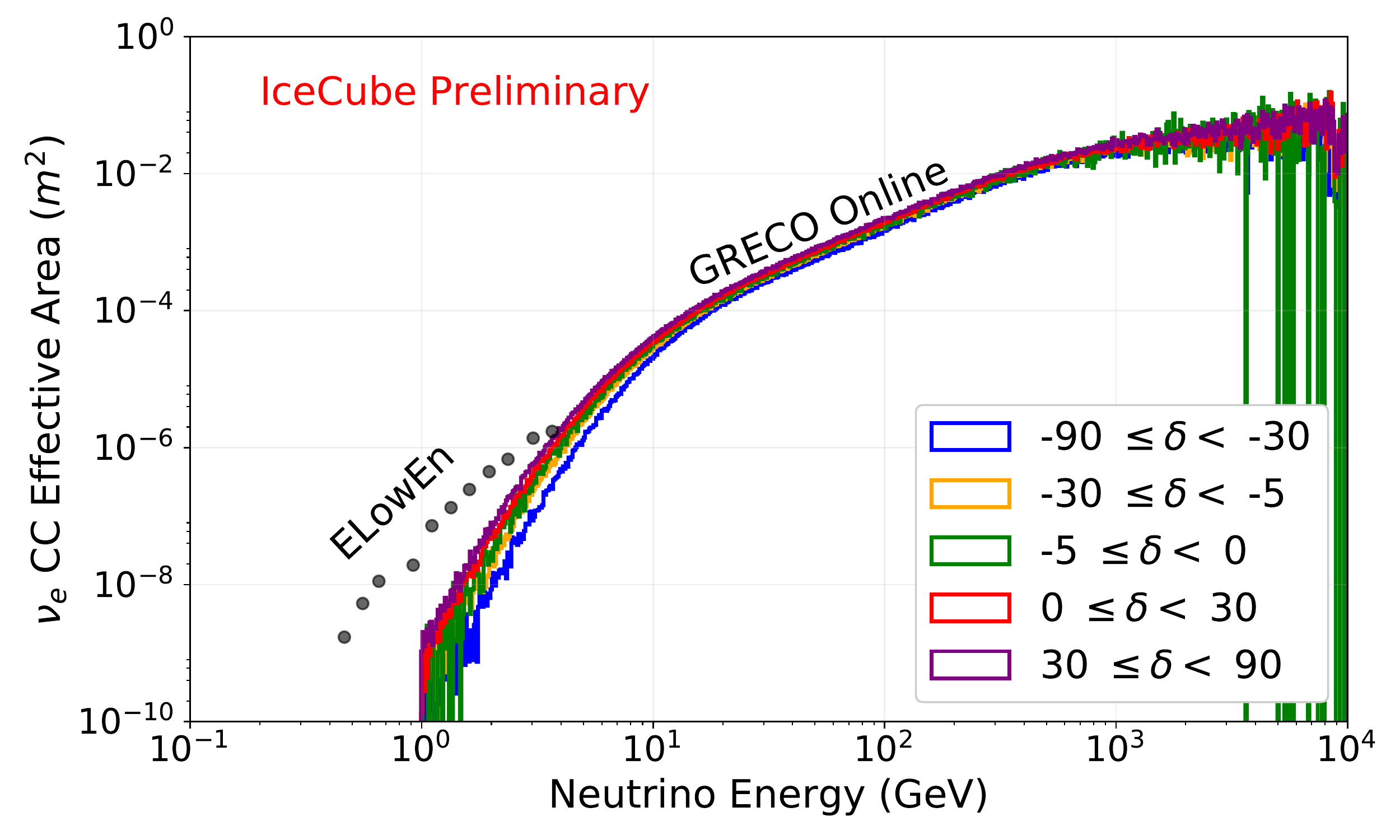}}
    \caption{The effective areas of the new GRECO Astronomy sample for charged-current muon neutrino interactions (left) and charged-current electron neutrino interactions (right). The through-going muon neutrino gamma-ray followup (GFU) selection\cite{Aartsen:2016qbu} and the all-flavor extremely low energy (ELowEn) selection\cite{deWasseige:2019xcl} are shown for comparison.}
    \label{fig:greco_astro_aeff}
\end{figure}

The new selection, referred to as the GRECO Astronomy sample, provides significantly improved effective areas relative to the GRECO oscillation sample at energies below 100 GeV and is competitive with higher energy through-going track selections up to several TeV, as shown in Figure~\ref{fig:greco_astro_aeff}.

As with the GRECO oscillation selection, the new GRECO Astronomy selection provides a nearly flavor- and declination-independent significant effective area improvement in the 1-100 GeV range.
The GRECO Astronomy selection also significantly boosts the effective area available for searches in the southern sky relative to existing through-going track samples, allowing GRECO to improve southern sky limits up to tens of TeV.

\begin{figure}
    \centering
    \subfloat{\includegraphics[height=3.5cm, trim=0 0 135 0, clip]{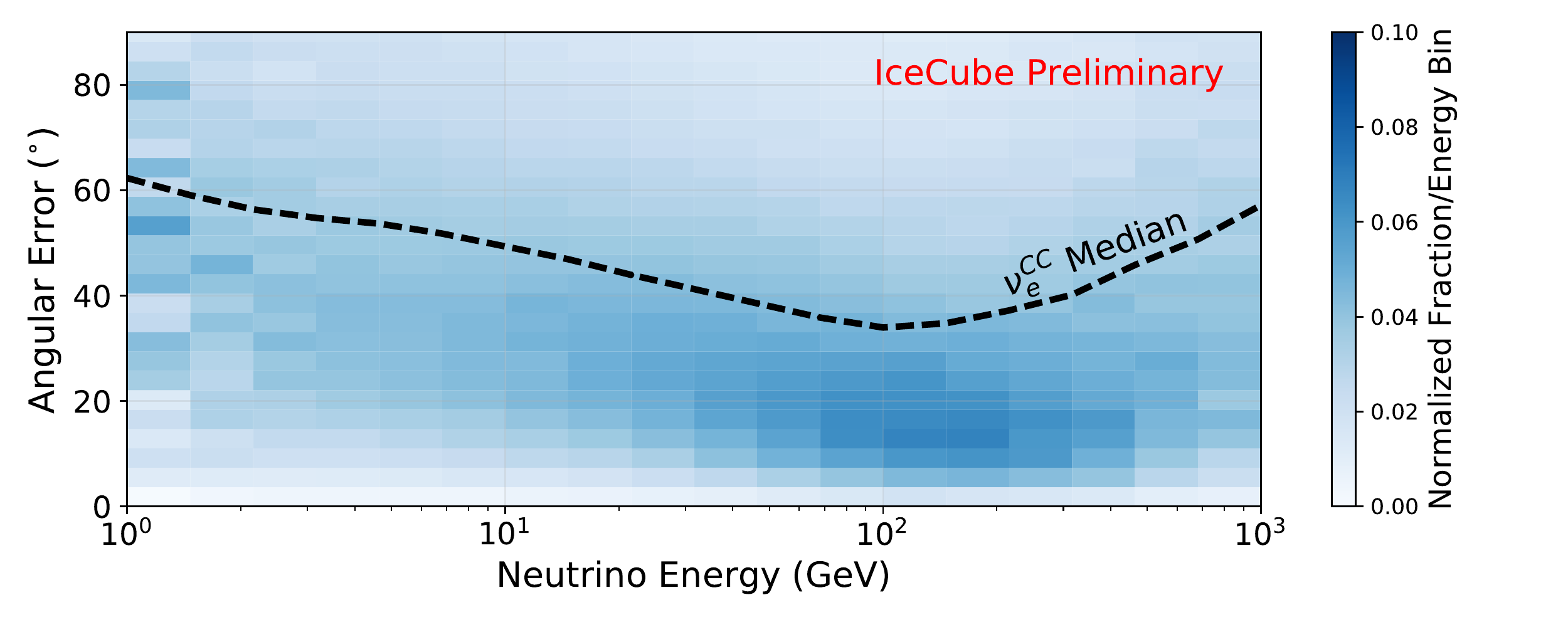}}
    \subfloat{\includegraphics[height=3.5cm, trim=0 0 50 0, clip]{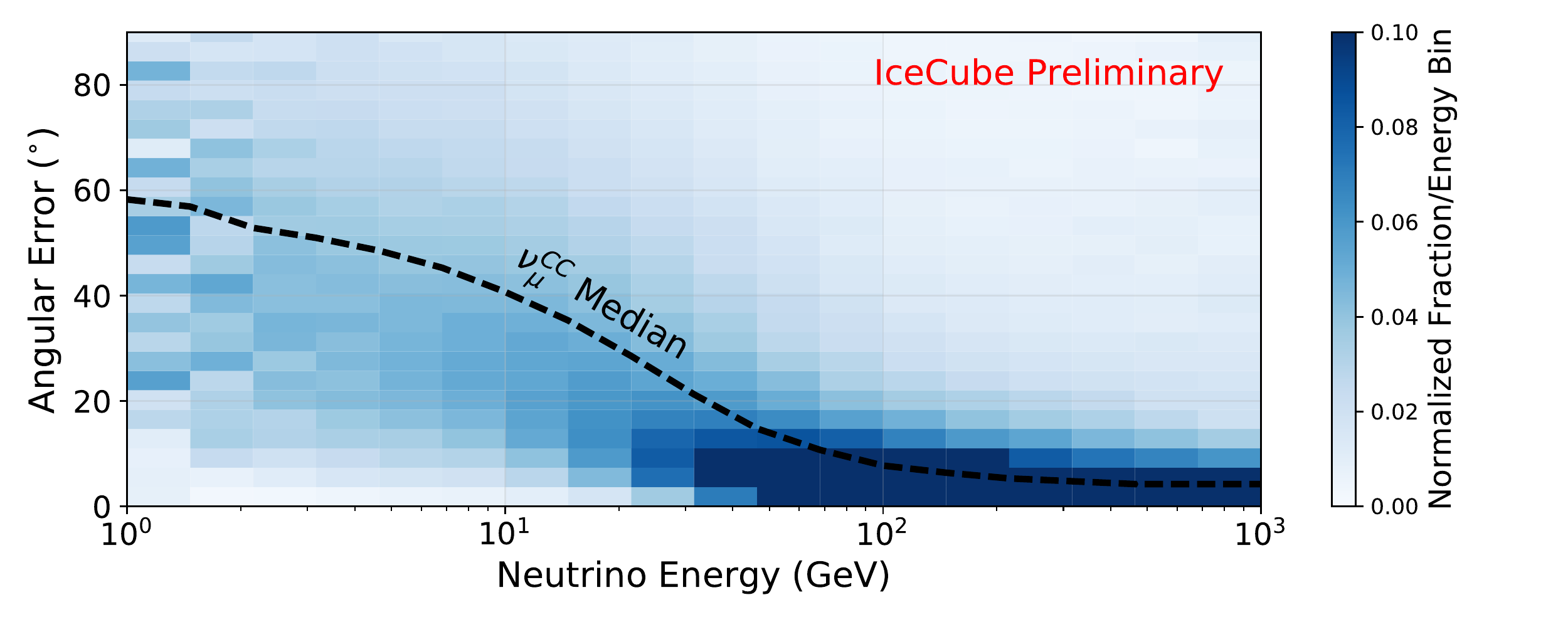}}
    \caption{Reconstruction performance in the GRECO astronomy sample for simulated charged-current electron (left) and muon (right) neutrinos as a function of neutrino energy.}
    \label{fig:greco_astro_resol}
\end{figure}
Reconstructions of events in the GRECO astronomy sample are performed using the same algorithm as in the GRECO oscillation sample.
Interactions with outgoing muons yield average reconstruction angular errors of 5-10$^\circ$ above 100 GeV, but events at lower energies tend to be poorly reconstructed, with average errors up to 40$^\circ$. 
The distribution of angular errors for electron and muon neutrinos are shown in Figure~\ref{fig:greco_astro_resol}.
Angular error estimates $\sigma_i$ are produced for each event using a dedicated BDT trained to estimate the angular distance between reconstructed direction and the original neutrino direction.

\section{Searches for Novae with GRECO Online}
\label{sec:novae}
The GRECO Astronomy sample has potential to search not only for untriggered neutrino flares, as is done in the GRECO oscillations sample analysis described in Section~\ref{sec:mia_analysis}, but also to constrain local low energy neutrino sources. A prime candidate for this type of search is Galactic Novae, the luminous outbursts that occur when a white dwarf in a binary system rapidly accretes matter from its companion star, leading to unstable nuclear burning on the surface of the white dwarf. It was recently discovered that novae, typically identified in optical wavelengths, were often accompanied by GeV gamma rays \citep{Abdo:2010he}. To date, the Large Area Telescope (LAT) aboard NASA's \textit{Fermi} satellite, has detected over one dozen novae. For a recent review of novae, see \citep{Chomiuk:2020zek}.

\begin{figure}
    \centering
    \includegraphics[width=\textwidth]{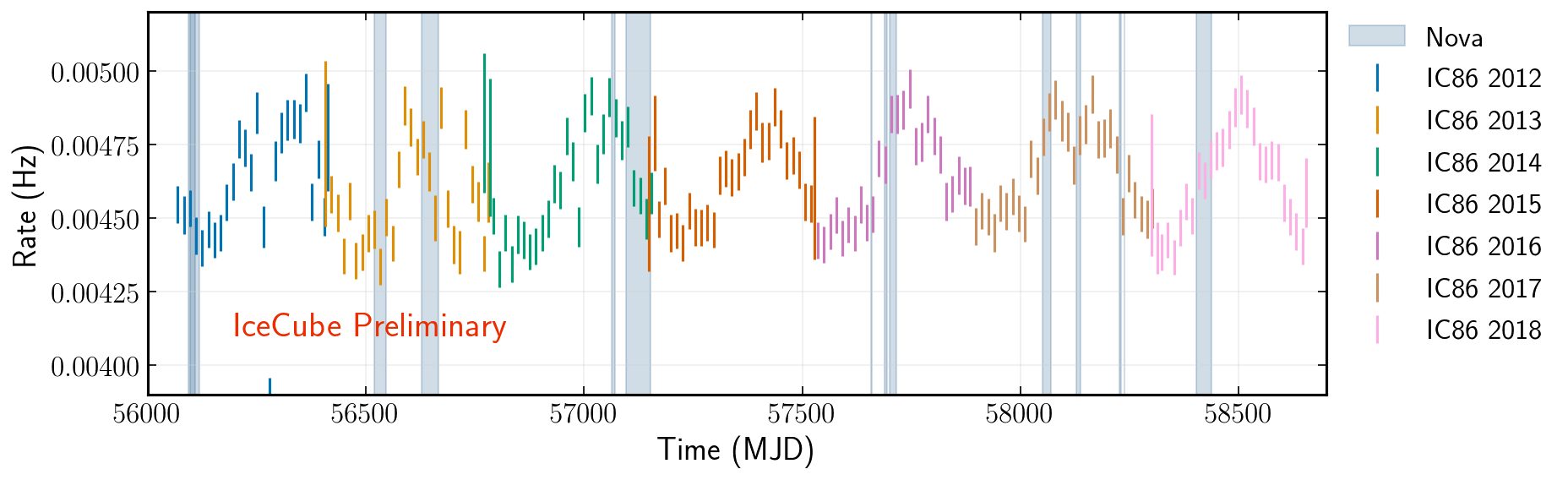}
    \caption{Rates of the GRECO astronomy selection over time. Seasonal fluctuations from the atmospheric backgrounds are visible around the average rate of 4.5 mHz. The times of catalog novae occurring during the avaible time periods of the GRECO astronomy selection are included.}
    \label{fig:greco_astro_rates}
\end{figure}

The gamma-ray emission from novae was recently found to be strongly correlated in time with the optical emission, lending evidence for a common origin in shocks \citep{Aydi:2020znu}. These non-thermal gamma rays could, in principle, be produced from a variety of either leptonic or hadronic mechanisms, depending on the composition of the relativistic particles accelerated at the shocks. 
While several arguments currently favor the hadronic scenario,
neutrinos could hold the key to distinguishing between the leptonic and hadronic models and could provide valuable information for understanding the environments of these shocks. However, as novae are believed to have a maximum acceleration threshold up to only around TeV energies for initial cosmic-ray primaries, traditional track-based neutrino searches are not sensitive to signals from novae, but the GRECO astronomy sample can be used to search for these signals\cite{Pizzuto2020Search}. 

To search for such a signal, we use a similar likelihood as defined in Eq.~\ref{eq:likelihood}, but the signal and background PDFs, $\mathcal{S}_i$ and $\mathcal{B}_i$, are functions not only of directional observables, but also of the reconstructed energies of the neutrino candidate events. A total of 16 novae previously identified by Fermi-LAT (including candidate gamma-emitters from \cite{Franckowiak:2017iwj}) that are coincident with the GRECO astronomy livetime, shown in Figure~\ref{fig:greco_astro_rates},  are used as a catalog. As with the analysis described in Section~\ref{sec:mia_analysis}, for large time-windows, the atmospheric backgrounds become overwhelming, and the search must be limited to shorter timescales. Although some novae are seen to emit for weeks to months, preliminary studies suggest a maximum allowed integration time of $\sim 10$ days for this analysis.

Working under the assumption that the detected gamma rays from novae are completely produced from the decay of neutral pions, and neglecting any gamma-ray attenuation, expected signals can be estimated using the procedure outlined in \cite{Halzen:2018iak}. Using the best-fit single power law spectra for each of these novae yield preliminary median 90\% sensitivities
% – defined here as the median 90\% confidence level upper limit that would be placed under the assumption of a background-only hypothesis – 
to individual novae at the level of about $\sim 10^1 - 10^2$~GeV~cm$^{-2}$, when constraining the energy-scaled time-integrated flux, $E^2 dN/dE \Delta T$ at 1 GeV, and when using an analysis window of one day. While the expectation from each individual nova is at least one order of magnitude below this, the technique remains sensitive to nearby or exceptionally bright novae, and it also neglects any gamma-ray attenuation, which would increase the relative neutrino to gamma-ray flux. Additionally, stacking together the signals from the entire catalog of novae, or searching for ways to remain sensitive to longer timescales, would improve the capabilities of this analysis. Both of these improvements are currently underway.

\bibliographystyle{ICRC}
\bibliography{references}

% Full authors list (ONLY FOR COLLABORATIONS)
\clearpage
\section*{Full Author List: IceCube Collaboration}

% \noindent \textbf{Note comment afterwards:} Collaborations have the possibility to provide an authors list in xml format which will be used while generating the DOI entries making the full authors list searchable in databases like Inspire HEP. For instructions please go to icrc2021.desy.de/proceedings or contact us under icrc2021proc@desy.de.\\

% \scriptsize
% \noindent
% first.author$^1$, 
% second.author$^2$, 
% third.author$^3$ % .... more names
% and 
% last.author$^{n}$ \\

% \noindent
% $^1$first.affiliation.
% $^2$second.affiliation. % .... more affiliation
% $^{m}$last.affiliation.

\scriptsize
\noindent
R. Abbasi$^{17}$,
M. Ackermann$^{59}$,
J. Adams$^{18}$,
J. A. Aguilar$^{12}$,
M. Ahlers$^{22}$,
M. Ahrens$^{50}$,
C. Alispach$^{28}$,
A. A. Alves Jr.$^{31}$,
N. M. Amin$^{42}$,
R. An$^{14}$
% ...
\\

\noindent
% ...
$^{12}$ Universit{\'e} Libre de Bruxelles, Science Faculty CP230, B-1050 Brussels, Belgium \\
% ...
$^{14}$ Department of Physics and Laboratory for Particle Physics and Cosmology, Harvard University, Cambridge, MA 02138, USA \\
% ...
$^{17}$ Department of Physics, Loyola University Chicago, Chicago, IL 60660, USA \\
$^{18}$ Dept. of Physics and Astronomy, University of Canterbury, Private Bag 4800, Christchurch, New Zealand \\
% ...
$^{22}$ Niels Bohr Institute, University of Copenhagen, DK-2100 Copenhagen, Denmark \\
% ...
$^{28}$ D{\'e}partement de physique nucl{\'e}aire et corpusculaire, Universit{\'e} de Gen{\`e}ve, CH-1211 Gen{\`e}ve, Switzerland \\
% ...
$^{31}$ Karlsruhe Institute of Technology, Institute for Astroparticle Physics, D-76021 Karlsruhe, Germany  \\
% ...
$^{42}$ Bartol Research Institute and Dept. of Physics and Astronomy, University of Delaware, Newark, DE 19716, USA \\
% ...
$^{50}$ Oskar Klein Centre and Dept. of Physics, Stockholm University, SE-10691 Stockholm, Sweden \\
% ...
$^{59}$ DESY, D-15738 Zeuthen, Germany \\
% ...

\subsection*{Acknowledgements}

\noindent
USA {\textendash} U.S. National Science Foundation-Office of Polar Programs,
U.S. National Science Foundation-Physics Division,
U.S. National Science Foundation-EPSCoR,
Wisconsin Alumni Research Foundation,
Center for High Throughput Computing (CHTC) at the University of Wisconsin{\textendash}Madison,
Open Science Grid (OSG),
Extreme Science and Engineering Discovery Environment (XSEDE),
Frontera computing project at the Texas Advanced Computing Center,
U.S. Department of Energy-National Energy Research Scientific Computing Center,
Particle astrophysics research computing center at the University of Maryland,
Institute for Cyber-Enabled Research at Michigan State University,
and Astroparticle physics computational facility at Marquette University;
Belgium {\textendash} Funds for Scientific Research (FRS-FNRS and FWO),
FWO Odysseus and Big Science programmes,
and Belgian Federal Science Policy Office (Belspo);
Germany {\textendash} Bundesministerium f{\"u}r Bildung und Forschung (BMBF),
Deutsche Forschungsgemeinschaft (DFG),
Helmholtz Alliance for Astroparticle Physics (HAP),
Initiative and Networking Fund of the Helmholtz Association,
Deutsches Elektronen Synchrotron (DESY),
and High Performance Computing cluster of the RWTH Aachen;
Sweden {\textendash} Swedish Research Council,
Swedish Polar Research Secretariat,
Swedish National Infrastructure for Computing (SNIC),
and Knut and Alice Wallenberg Foundation;
Australia {\textendash} Australian Research Council;
Canada {\textendash} Natural Sciences and Engineering Research Council of Canada,
Calcul Qu{\'e}bec, Compute Ontario, Canada Foundation for Innovation, WestGrid, and Compute Canada;
Denmark {\textendash} Villum Fonden and Carlsberg Foundation;
New Zealand {\textendash} Marsden Fund;
Japan {\textendash} Japan Society for Promotion of Science (JSPS)
and Institute for Global Prominent Research (IGPR) of Chiba University;
Korea {\textendash} National Research Foundation of Korea (NRF);
Switzerland {\textendash} Swiss National Science Foundation (SNSF);
United Kingdom {\textendash} Department of Physics, University of Oxford.

\end{document}